\documentclass[a4paper,11pt]{article}

\usepackage{jheppub} 

\def\m#1{\mathrm{#1}} 
\def\be#1{\begin{equation}#1\end{equation}} 
\def\beqnn#1{\begin{eqnarray}#1\end{eqnarray}}
\def\integral#1#2{\int^{#2}_{#1}} 
\def\summation#1#2{\sum^{#2}_{#1}} 
\def\product#1#2{\prod^{#2}_{#1}} 
\def\production#1#2{\prod^{#2}_{#1}} 

\title{\boldmath 
Direct computational approach to lattice supersymmetric quantum mechanics}

\author[a]{Daisuke Kadoh, \note{}}
\author[b,c]{Katsumasa Nakayama}

\affiliation[a]{Research and Educational Center for Natural Sciences, Keio University, \\ Yokohama 223-8521, Japan}
\affiliation[b]{Department of Physics, Nagoya University, \\ Nagoya, 464-8602, Japan}
\affiliation[c]{KEK Theory Center, High Energy Accelerator Research Organization (KEK), Tsukuba 305-0801, Japan}

\emailAdd{kadoh@keio.jp}
\emailAdd{katumasa@post.kek.jp}

\abstract{
We propose a numerical method of estimating various physical quantities 
in lattice (supersymmetric) quantum mechanics.
The method consists only of deterministic processes such as computing a product of transfer matrix, 
and has no statistical uncertainties.
We use the numerical quadrature to define the transfer matrix as a finite dimensional matrix, 
and find that it effectively works by rescaling variable for sufficiently small lattice spacings. 
For a lattice supersymmetric quantum mechanics, 
the correlators can be estimated without statistical errors, and  
the effective masses coincide with the exact solution within very small errors less than $0.001$\%. 
The SUSY Ward identity is also precisely studied in compared with the Monte-Carlo method. 
Our method is not limited to a lattice SUSY quantum mechanics, but is also applicable to any other lattice models of quantum mechanics.

}

\begin{document}
\maketitle
\flushbottom

\section{Introduction}
\label{sec:introduction}

Supersymmetric quantum mechanics (SUSY QM)  has been  extensively studied  
since it provides a good testing ground for supersymmetric field theories \cite{Witten:1981nf,Witten:1982df,Cooper:1994eh}.
Supersymmetry breaking triggered by instantons has been deeply understood 
and the associated  topological index  (Witten index) are widely used in other theories. 
In early studies, 
a class of exactly solvable potentials, so-called shape invariant potentials (SIP), 
was discovered and has attracted much attentions in revealing various aspects of SUSY QM \cite{Gendenshtein:1984vs}.   
However, to learn SUSY models beyond those cases, 
numerical approaches including lattice theory could play a significant role.

There have been many attempts to define SUSY QM and low dimensional Wess-Zumino model on the lattice 
\cite{
Dondi:1976tx,
Elitzur:1982vh,
Elitzur:1983nj,
Cecotti:1982ad, 
Sakai:1983dg,
Golterman:1988ta, 
Kikukawa:2002as,
Giedt:2004qs,
Feo:2004kx,
Kato:2008sp,
Kadoh:2009sp,
Kadoh:2010ca,
Kato:2013sba,
Kadoh:2015zza, 
Asaka:2016cxm,
Kato:2016fpg,
DAdda:2017bzo}. 
The proposed lattice models have provided not only further understanding of SUSY models 
but also testing grounds for lattice formulations of higher dimensional SUSY theories
\cite{
Beccaria:1998vi,
Catterall:2000rv,
Catterall:2001fr,
Giedt:2004vb,
Giedt:2005ae,
Bergner:2007pu,
Kanamori:2007yx,
Kastner:2008zc,
Bergner:2009vg,
Kawai:2010yj, 
Kanamori:2010gw,
Wozar:2011gu,
Schierenberg:2012pb,
Steinhauer:2014yaa}.
As for numerical studies, 
the standard Monte-Carlo techniques have been utilized so far in many literatures.
It is, however, a demanding task to obtain precise results from Monte-Carlo simulations,
especially for the case of supersymmetry breaking that causes the notorious sign problem.

Recently, a direct computational method has been proposed
by Baumgartner and Wenger \cite{Baumgartner:2014nka,Baumgartner:2015qba,Baumgartner:2015zna}.
This approach could help us to solve
the sign problem because a product of transfer matrix is directly estimated 
without any stochastic processes in principle.
In this sense, it is similar to the tensor network renormalization (TNR) \cite{Levin:2006jai}  which has developed over the last decade,
including applications to field theory 
\cite{
Shimizu:2012wfa,
Shimizu:2014uva,
Shimizu:2014fsa,
Takeda:2014vwa,
Kawauchi:2016xng,
Sakai:2017jwp,
Yoshimura:2017jpk, 
Shimizu:2017onf,
Kadoh:2018hqq}. 
In TNR, the partition function and correlators are represented as a product of tensors, 
which are estimated by coarse graining of network.  
Those two methods share common idea in the sense that a product of local matrix or 
tensor is directly estimated by matrix operations.
Keeping this point in mind,  we may say that
further developments in SUSY QM using transfer matrix are associated with TNR 
and applications to higher dimensional field theories.

In this paper, we propose an explicit and simple computational method in lattice (SUSY) QM, which is 
based on the transfer matrix representation of Euclidean path integrals. 
Although the transfer matrix is an infinite dimensional matrix, we  express it as 
a finite dimensional ones by approximating the integrals of coordinate.
The well-known Gaussian quadrature is used for the  approximation.  
As  we will see later,  such simple quadrature effectively works 
with some scale transformation of variables.
In this procedure, we can estimate the correlators 
in tremendous lattice volume with sufficiently small lattice spacings.

We test our method for two cases without supersymmetry breaking, 
$\phi^6$ theory and a SIP potential (Scarf I). 
Although the energy spectra can be directly estimated from transfer matrices, 
we use the standard technique to evaluate the effective masses from correlators, 
which are used in the Monte-Carlo method, in order to compare our results with Monte-Carlo ones. 
The estimated masses reproduce the known results in high precision.   
For Scarf I potential, energy eigenvalues are exactly solvable, and the lowest one for $A/\alpha=10,B=0$ in (\ref{scarf}) is  
\begin{eqnarray}
m_{exact}/\alpha^2 = 10.5.
\end{eqnarray}
As the effective mass, we have
\begin{eqnarray}
m/\alpha^2  =10.49998(2).
\end{eqnarray}
Thus one can say that our approach effectively works compared with the previous Monte Carlo simulations.

This paper is organized as follow. 
In section \ref{sec:susyqm}, SUSY QM is given in the path integral formulation with Euclidean time,  
and a lattice theory which partially keeps supersymmetry is also given. 
We propose a direct computational method using the transfer matrix in section \ref{our_method}.
The numerical results for $\phi^6$-theory  and Scarf I potential are given in section \ref{results}.
We conclude this paper with some discussions in section \ref{sec:conclusion}.

\section{Supersymmetric quantum mechanics}
\label{sec:susyqm}

We start with reviewing ${\cal N}=2$ supersymmetric quantum mechanics which is defined 
in Euclidean path integral formulation. 
The lattice theory is then defined according to Refs.\cite{Catterall:2000rv} in  a form
which retains one supersymmetry exactly on the lattice.

\subsection{Continuum theory}
The supersymmetric quantum mechanics is described by  
a coordinate $\phi(t) \in \mathbb{R} $ and one-component Grassmann 
variables $\psi(t)$ and $\overline{\psi}(t)$.
We assume that they obey the periodic boundary condition with a period $\beta$
for the Euclidean time $t \in \mathbb{R}$.
%
%
The action of SUSY QM is given by 
\be{
\label{cont_action}
S
=
\integral{0}{\beta}\m{d}t
\left\{
\frac{1}{2}
(\partial_t \phi(t) )^2
+
\frac{1}{2}
W(\phi(t))^2
+
\overline{\psi}(t)
\big( \partial_t + W'(\phi (t))\big)
\psi (t)
\right\}.
}
We refer to $W(\phi)$, which is an arbitrary function of $\phi$, as a superpotential.
\footnote{
For $W=\frac{d }{ d \phi} h$, $h$ is often referred to as a superpotential in several literatures.
}  
For any superpotential, 
(\ref{cont_action}) is invariant under supersymmetry transformation, 
\beqnn{
\label{super1}
&&
\delta \phi = \epsilon \psi + \overline{\epsilon} \overline{\psi},
\\
&& 
\delta \psi = 
-\overline{\epsilon} (\partial_t \phi - W(\phi) ),
\\
&&
\label{super3}
\delta \overline{\psi} = -\epsilon (\partial_t \phi + W(\phi)),
}
where $\epsilon$ and $\overline{\epsilon}$ are one-component Grassmann numbers.



In the path integral formulation, the partition function is given by
\footnote{ The fermion measure is normalized such that $Z=1$ at large $m$ in the free theory. }
\be{
\label{z}
Z=\int D\phi D\overline{\psi} D\psi  \,{\rm e}^{-S},
} 
and  the correlation functions are also defined in the usual manner. 
For periodic boundary condition, $Z$ is just the Witten index ${\rm Tr} {(-1)^F}$
which measures the difference of the number of bosonic and fermionic vacuum states. 
In the case of SUSY QM, we can perfectly learn from the index whether supersymmetry is broken or not: 
$Z=0$ if and only if supersymmetry is broken. 
As well-known, this index is unchanged under any deformation of parameters 
which retains the asymptotic behavior of $W(\phi)$. In the case of $|W(\pm \infty)|=\infty $,  
supersymmetry is unbroken when the signs of $W(\infty)$ and $W(-\infty)$ are different from each other,
while supersymmetry is broken when they are the same \cite{Witten:1981nf,Witten:1982df}.

In this paper, we focus on two cases which do not exhibit supersymmetry breaking: 
$\phi^6$ theory and Scarf I potential.
The $\phi^6$ theory is given by the cubic superpotential,
\be{
\label{cubic_potential}
W(\phi) = m
\phi
+
g
\phi^3,
}
with  the mass $m$ and coupling $g >0$,
and  Scarf I potential is given by
\be{
\label{scarf}
W(\phi) = A {\rm tan} (\alpha \phi) - B{\rm sec}(\alpha \phi), \qquad  -\frac{\pi}{2} \le \alpha \phi \le \frac{\pi}{2},
}
where $A>B \ge 0$ and $\alpha > 0$.
We have $Z =1$ for these  two cases.  
In section \ref{results}, we will test our computational method proposed in section \ref{our_method} for 
(\ref{cubic_potential}) and (\ref{scarf}). 
The former potential is used to compare our results with the previous Monte-Carlo results 
since it has been studied in many literatures.    
The later one provides a comparison between our results and exact solution
because it is one of shape invariant potentials (SIPs) which are exactly solvable \cite{Cooper:1994eh}.

\subsection{Lattice theory}
\label{lat_theory}

The lattice theory is defined by considering the bosonic and fermionic coordinates as variables $\phi_t$ and $\psi_t$ living on the integer sites, $t \in \mathbb{Z}$.
We now take the lattice spacing $a=1$ without loss of generality, 
and often express $a$-dependence of dimensionful quantities explicitly.
We again assume that all variables $\phi_t$, $\psi_t$, $\overline{\psi}_t$ satisfy the periodic boundary 
condition of the period $\beta=Na$, where $N \in \mathbb{N}$ is the lattice size.


The lattice action is given by 
\beqnn{
\label{lat_action}
S
&=&
\summation{t=1}{N}
\left[
\frac{1}{2}
(\nabla \phi_t
+
 W(\phi_t))^2
+
\overline{\psi}_t
(\nabla + W'(\phi_t))
\psi_t
\right], 
}
where the backward difference operator,
\be{
\nabla\phi_t = \phi_t - \phi_{t-1}}
is used to approximate the derivative $\partial_t$. 
More generally, we can use $\nabla=\frac{\nabla_+  + \nabla_-}{2} - \frac{r}{2} \nabla_+ \nabla_-$
with forward and backward difference operators 
$\nabla_+ $, $\nabla_-$, and the Wilson parameter $r$. 
Here we have the backward one  by setting $r=1$ for simplicity. 

We now consider lattice supersymmetry transformation which  
is defined by replacing $\partial_t$ in (\ref{super1})-(\ref{super3}) by $\nabla_\pm$ as
\beqnn{
&&
\delta \phi = \epsilon \psi + \overline{\epsilon} \overline{\psi},
\\
&& 
\delta \psi = 
-\overline{\epsilon} (\nabla_+ \phi - W(\phi) ),
\\
&&
\delta \overline{\psi} = -\epsilon (\nabla_- \phi + W(\phi)).
}
For free theory, the lattice action (\ref{lat_action})  is invariant under this transformation. 
For interacting theories, it is, however, not invariant for any $\epsilon$ and $\overline{\epsilon}$; 
one-supersymmetry ($\epsilon =0,\overline{\epsilon} \neq 0)$ remains as an exact symmetry, 
while the other  ($\epsilon \neq 0,\overline{\epsilon} =0)$ is broken at finite lattice spacing.
The full ${\cal N}=2$ supersymmetry is shown 
to be restored in the continuum limit as seen 
in Ref\cite{Catterall:2000rv,Giedt:2004vb,Bergner:2007pu,Bergner:2009vg,Schierenberg:2012pb} 
and also precisely shown in section \ref{results}.

The partition function is given by (\ref{z}), and after integrating fermionic variables, we have
\be{
\label{lattice_z}
Z= \int D \phi \, e^{-S_B} {\rm det}(\nabla+W'(\phi)),
}
with 
\be{
\label{lat_boson_action}
S_B= 
\sum_{t=1}^{N}
\frac{1}{2}
(\nabla \phi_t
+
 W(\phi_t))^2.
}
The fermion determinant is well-defined since the size of matrix $\nabla+W'(\phi)$ is finite. 
In this case, we can explicitly write it down as
\footnote{
The first term of RHS is affected from the boundary condition.
In the case of anti-periodic boundary condition, $-1$ is replaced by $+1$ as
$\m{det}
(\nabla + W'(\phi))
=
1
+
\product{t=1}{N}
\left[
1
+
W'(\phi_t)
\right]$.
}
\be{
\m{det}
(\nabla + W'(\phi))
=
-1
+
\product{t=1}{N}
\left[
1
+
W'(\phi_t)
\right],
\label{fermi_det}
}
which is strictly positive for (\ref{cubic_potential}) and (\ref{scarf}). 
The integrations of $\phi $ are given by the finite dimensional integrals on the lattice as 
\be{
\label{boson_mearure}
\int D\phi \equiv \prod_{t=1}^N \int_{-\infty}^{\infty} \frac{d\phi_t}{\sqrt{2\pi}}.
}
Thus the Monte-Carlo techniques can be applied to this lattice model.


As seen in the following sections, 
the transfer matrix representation of (\ref{lattice_z})  is very useful to study this model. 
Using (\ref{fermi_det}) and (\ref{boson_mearure}), we can easily show that  
\be{
\label{z_t_rep}
Z=\int^{\infty}_{-\infty} d {\phi_1} d \phi_2 \ldots d \phi_N
\bigg\{
\prod_{t=1}^N
S_{\phi_t \phi_{t-1}}
-
\prod_{t=1}^N
T_{\phi_t \phi_{t-1}}
\bigg\}
}
where
\beqnn{
&& 
T_{\phi\phi'} = \frac{1}{\sqrt{2\pi}} \, {\rm exp} \left\{-\frac{1}{2} \left(\phi-\phi' + W(\phi) \right)^2 \right\}, 
\\
&& 
S_{\phi\phi'} = (1+W^\prime (\phi)) T_{\phi\phi'}.
}
 (\ref{z_t_rep}) is formally denoted as
\be{
\label{zts}
Z = {\rm Tr} (S^N) - {\rm Tr}(T^N),
}
where 
${\rm Tr}(A)=\int^\infty_{-\infty} d \phi A_{\phi\phi}$.
We now note that numerical approaches are not applicable to this representation 
in a straightforward way 
since the traces in (\ref{zts}) can not be performed for the infinite dimensional matrices
$T_{\phi\phi'}$ and $S_{\phi\phi'}$ with $\phi,  \phi' \in \mathbb{R}$.

The two transfer matrices $S$ and $T$ have specific meanings.
In SUSY QM, there are two Hamiltonian $H_B$ and $H_F$ 
which act on the fermionic and bosonic states, respectively\cite{Cooper:1994eh}.
Let $E^{B}_i (i=0,1,2,\ldots)$ be the eigenvalues of $H_B$ and let $E^{F}_i(i=1,2,\ldots)$ be ones of $H_F$ 
where $E^{B}_i$ and $E^F_i$ are sorted in ascending order. 
In the continuum theory, one can show that 
\be{
\label{EBandEF}
 E^{B}_0=0, \qquad \quad E^B_i=E^F_i >0 \quad (i=1,2,\cdots).
}
The transfer matrices are interpreted as  $S={\rm exp}(-aH_B)$ and $T={\rm exp}(-aH_F)$ on the lattice 
as discussed in Ref \cite{Baumgartner:2014nka,Baumgartner:2015qba,Baumgartner:2015zna}.
Then, we can estimate the energy spectra for bosonic and fermionic states 
from the eigenvalues of $S$ and $T$.

\section{Simple computational method of lattice (SUSY) QM}
\label{our_method}

\subsection{Partition function}

We first consider the partition function which is the simplest example 
to describe our computational method.

In order to express the transfer matrices as finite dimensional matrices, 
we replace  (\ref{boson_mearure}) by a numerical quadrature which is given as a weighted summation,
\footnote{The similar techniques are used in the TNR formulations for scalar systems
\cite{Shimizu:2012wfa, Kadoh:2018hqq}.}
\be{
\label{formal_quad}
\int^{\infty}_{-\infty} d\phi F(\phi) \approx 
\sum_{\phi \in S_K} g_K(\phi) F(\phi),
}
where $g_K(\phi)$ is a weight function and 
$S_K$ is a set of $K$ points which are used for the approximation.
$g_K$ and $S_K$ is determined for each quadrature rule.
For instance, 
the Gauss-Hermite quadrature is provided by
\be{
g_K(\phi)= \frac{2^{K-1} K! \sqrt{\pi}}{K^2 H_{K-1}(\phi)} e^{\phi^2}, 
}
and taking $S_K$ as a set of roots of $K$-th Hermite polynomial $H_K$.
We expect that in (\ref{formal_quad}) the true value of LHS can be obtained within small systematic errors 
by evaluating RHS 
at sufficiently large $K$ 
for well-behaved function $F(\phi)$ for which the quadrature effectively works.

The partition function (\ref{lattice_z}) is now approximated by replacing each measure of $\phi$ by (\ref{formal_quad}), 
and (\ref{z_t_rep}) turns out to be 
\be{
\label{zt0}
Z \approx \sum_{\phi_1 \in S_K}   \sum_{\phi_2 \in S_K}  
\ldots  \sum_{\phi_N \in S_K}  
\bigg\{
\prod_{t=1}^N
S_{\phi_t \phi_{t-1}}
-
\prod_{t=1}^N
T_{\phi_t \phi_{t-1}}
\bigg\}
}
where
 \beqnn{
&& 
\label{naiveT}
T_{\phi\phi'} = \sqrt{\frac{g_K(\phi) g_K(\phi')}{2\pi}} \, {\rm exp} \left\{-\frac{1}{2} \left(\phi-\phi' + W(\phi) \right)^2 \right\}, 
\\
&& 
\label{naiveS}
S_{\phi\phi'} = (1+W^\prime (\phi) ) T_{\phi\phi'}, 
}
Since $T$ and $S$ are matrices of size $K$, 
(\ref{zt0}) can be written as
\be{
\label{zt}
Z \approx  {\rm tr} (S^N) - {\rm tr}(T^N),
}
where $"{\rm tr}"$ is the trace of $K$ by $K$ matrices. 
Note that $T$ and $S$ are not uniquely determined 
because (\ref{zt}) is invariant under similarity transformations of them.

We can estimate $Z$ by computing the RHS of (\ref{zt}) at large $K$ once an effective quadrature is found.
One might think that conventional methods 
such as trapezoidal rule, Simpson's rule, or a kind of Gaussian quadrature
are less effective for multiple integrations 
since the number of required sample points increases exponentially with multiplicity. 
Besides, (\ref{zt0}) is likely ineffective 
since it is just a superposition of one-dimensional quadrature (\ref{formal_quad}).
However, numerical results shown in section \ref{results} show that 
a simple Gaussian quadrature effectively works 
by using it with a scale transformation of variables, 
explained in section \ref{improvement}.

\subsection{Correlation functions}
\label{correlator}

The correlation functions are estimated in a similar manner as with the partition function. 
As simple examples, we present two-point functions of 
$\phi_t$ and $\psi_t$. 
It is very easy to extend our method to general correlation functions 
which are given as a product of local operators ${\cal O}_i(t)$.\footnote{We assume that ${\cal O}_i(t)$ consists only of 
$\phi_t, \psi_t$ and $\phi_{t-1}, \psi_{t-1}$.
} 

The expectation value of an operator ${\cal O}$ is defined by
\be{
\langle {\cal O} \rangle
 = \frac{ \langle \langle
{\cal O}  \rangle \rangle}{Z},
}
where 
\be{
\label{numerator}
\langle \langle
{\cal O}  \rangle \rangle \equiv 
\int D\phi D  \overline{\psi} D \psi \, {\cal O} \, e^{-S}.
}
For later use, let us define the fermionic part of (\ref{numerator}) as
\be{
\langle \langle
{\cal O}  \rangle \rangle_F = \int D \overline{\psi} D\psi \, {\cal O} \, e^{-S_F},
}
with 
\be{
\label{lattice_fermi_action}
S_F=\sum_{t=1}^N 
\left\{ (1 + W'(\phi_t))  \overline{\psi}_t \psi_t 
- \overline{\psi}_{t} \psi_{t-1}
\right\}.
}
%
In our approach, $Z$ and $\langle \langle {\cal O}  \rangle \rangle$ are individually estimated 
in contrast to the Monte-Carlo method which provides the expectation value $\langle {\cal O} \rangle$ itself. 

We can easily show that $\langle \langle
\phi_t \phi_s  \rangle \rangle$ is written as a product of matrices 
by expressing the operator insertion as impurity matrix $D_{\phi \phi'} = \phi \cdot \delta_{\phi \phi'}$: 
\be{
\label{phiphi}
\langle \langle \phi_t \phi_s  \rangle \rangle 
= {\rm tr}\left( D S^{N-k}  D S^{k} \right)
-{\rm tr}\left( D T^{N-k}  D T^{k} \right), 
}
where $k=(s-t) \, {\rm mod} \,N$. \footnote{
$x \, {\rm mod} \, N$ is basically the reminder that is obtained dividing $x$ by $N$, 
and here we use $kN \, {\rm mod} \, N = N \, {\rm for}\, k \in \mathbb{Z}$.
}
The translational invariance and $ \langle \phi_t \phi_s \rangle= \langle \phi_s \phi_t \rangle $  
are manifestly follows from this representation. 
(\ref{phiphi}) is extended to a case of
two local operators made of $\phi_t$ and $\phi_{t-1}$,
${\cal O}_i(t)= \tilde {\cal O}_i[\phi_t,\phi_{t-1}]$:  
\be{
\label{oo}
\langle \langle {\cal O}_1(t) {\cal O}_2(s)  \rangle \rangle 
= {\rm tr}\left( \tilde {\cal O}_1 S^{N-k}   \tilde {\cal O}_2 S^{k} \right)
-{\rm tr}\left( \tilde {\cal O}_1 T^{N-k}   \tilde {\cal O}_2 T^{k} \right),
}
where $k=(s-t) \, {\rm mod} \,N$ and $\tilde {\cal O}_i[\phi,\phi']$ are regarded as a matrix 
with the column $\phi$ and the row $\phi'$.

Similarly, $\langle \psi_t  \overline{\psi}_s\rangle$ is given in terms of a product of the transfer matrices.
Since $\psi_t$ is connected to $ \overline{\psi}_s$ by the hopping term of (\ref{lattice_fermi_action}),   
it is easy to show that, for $k=1,\cdots,N$,
\beqnn{
&&
\langle \langle \psi_N \overline{\psi}_k \rangle \rangle_F = 
\int
d\overline{\psi}_{k-1}
d \psi_{k-1}
\ldots
d \overline{\psi}_{1}  
d {\psi}_{1}
\left\{
\prod_{t=1}^{k-1} e^{ -(1 + W'(\phi_t))  \overline{\psi}_t \psi_t 
}
\right\}
\nonumber 
\\
&& \hspace{2cm}
\times
\int 
d\overline{\psi}_N
d \psi_N
\ldots 
d \overline{\psi}_{k} 
d {\psi}_{k}
 \left\{
\prod_{t=k+1}^{N} e^{ \overline{\psi}_t \psi_{t-1} }
\right\}
\psi_N  \overline{\psi}_k,
}
and integrating fermionic variables, 
\be{
\label{psipsi_f}
\langle
\langle\psi_N\overline{\psi_{k}}\rangle \rangle_F
=
\production{t=1}{k-1}
\left(
1+W'(\phi_t)
\right).
}
Instead of (\ref{fermi_det}), we have (\ref{psipsi_f}) as a contribution of the fermion-part.
Finally we find that
\be{
\label{psipsi}
\langle\langle\psi_t\overline{\psi_{s}}\rangle\rangle
=
\m{tr}
\left[
S^{k-1}T^{N-k+1}
\right],
}
where  $k=(s-t) \, {\rm mod} \,N$. 

Once the matrix representations like (\ref{phiphi}) and (\ref{psipsi}) are obtained, one can evaluate the expectation value 
directly by computing the matrix products 
and traces and dividing them by $Z$.

\subsection{Some improvements}
\label{improvement}

The efficiency of our method depends on whether the quadrature 
is compatible with the lattice action, 
{\it e.g.} $W$ and the difference operator $\nabla$. 
To improve the efficiency, let us consider a rescaling of $\phi$ in (\ref{lattice_z}) 
as 
\be{
\Phi_t = s \phi_t\, \qquad (s \neq 0).
} 
Approximating the integrals of $\Phi$ by a quadrature,
the same formula (\ref{zt0}) 
is derived for the modified transfer matrices,
 \beqnn{
\label{modified_TS}
&& 
T_{\Phi\Phi'} = \sqrt{\frac{g_K(\Phi) g_K(\Phi')}{2\pi |s|}} \, {\rm exp} \left\{-\frac{1}{2} \left(\frac{\Phi-\Phi'}{s} + W\left(\Phi/s \right) \right)^2 \right\}, 
\\
&& 
S_{\Phi\Phi'} = (1+W^\prime (\Phi/s) ) T_{\Phi\Phi'}.
}
In the free theory, Gauss-Hermite quadrature could be effective for large masses $ma \gg 1$ 
by taking $s=ma/\sqrt{2}$ so that the damping factor is just a Gaussian with rate one, $e^{-\Phi^2}$. 
Near the continuum limit $ma \ll 1$, one might think that such rescaling is less effective 
because $T$ and $S$ have long tails in the field space, which corresponds to the zero kinetic term, $\Phi=\Phi'$.
 For interacting cases, the situation becomes more complicated 
since we have to find an effective quadrature for each interaction term.
 
The Gauss-Hermite quadrature with rescaling $\phi$, however, effectively works by tuning $s$
even for interacting cases, as seen in the numerical results in section \ref{results}.
We can choose $s$ in such  a way that some exact relations are obtained with as high precision as possible.  
In section  \ref{results}, $s$ is determined such that $|Z-1|$ is minimized for each lattice spacing 
since the Witten index is exactly one thanks to exact supersymmetry. 
Then we find that the energy eigenvalues and correlators are obtained in high precision even for small lattice spacings.

The computational cost of our method basically behaves as ${\cal O}(K^3\m{log}N)$ for 
the partition function, 
while it is multiplied by $N^{n-1}$ for $n$-point function.
The lattice size dependence follows from computing 
a matrix product $T^{2n} $ as $(T^n)^2$ for $n=1,2,\ldots$, recursively.
The diagonalization of transfer matrices often significantly reduces the cost. 
We show that (\ref{zt}) is expressed as
\be{
\label{eigenZ}
Z \approx \sum_{i=1}^K (\lambda_i)^N - \sum_{i=1}^K (\rho_i)^N,
}
where $\lambda_i$ and $\rho_i$ are eigenvalues of $S$ and $T$, respectively. 
Solving all eigenvalues of square matrices with size $K$ 
requires ${\cal O}(K^3)$  as the cost. We actually need 
a few eigenvalues for large $N$ if the lowest eigenvalues are isolated. 
Furthermore, for Gaussian quadrature,  $T$ and $S$ given as (\ref{naiveT}) and  (\ref{naiveS})
 are actually sparse matrices 
since $g_K(\phi) \simeq 0$ for $|\phi| \gg 1$.
In the case, the diagonalization is not a demanding task, and it is better to use (\ref{eigenZ})
in estimating (\ref{zt}) for large $N$ (small lattice spacings for fixed physical lattice size).
The cost of computing correlation functions reduces in the similar manner.

Although one-point functions are evaluated in the same way as correlation functions shown in section \ref{correlator}, 
we may use another way with the numerical derivative. 
For given local bosonic operator ${\cal O}(t)$, let us consider 
\be{
S' =S  + p \sum_{t=1}^N {\cal O}(t).
}
The partition function $Z'$ with the action $S'$ is also expressed as the form (\ref{zt})
with modified $T'$ and $S'$. We have
\be{
\langle {\cal O}(t) \rangle 
= -\frac{1}{N}\frac{\partial}{\partial p} {\rm log}(Z') \bigg|_{p=0},
}
and the one-point local function can be estimated without using an impurity matrix.  
Also in the case, the cost reduces by diagonalizing $S'$ and $T'$.

\section{Numerical results}
\label{results}

We have proposed a method of computing the partition function and correlators 
based on the transfer matrices in the previous section.
In this section, we demonstrate our method for $\phi^6$-theory 
and Scarf I potential for which supersymmetry is unbroken.

\subsection{$\phi^6$-theory}

Let us consider the $\phi^6$-interactions given by the cubic superpotential, 
\be{
W(\phi) = m \phi  + m^2 \lambda \phi^3,
}
where $m$ is the mass and $\lambda=g/m^2$ is the dimensionless coupling.  
Then, any dimensionful quantities are given in a unit of the physical scale $m$. 
For instance, $ma$ is the physical lattice spacing and $\beta m $ is the period (lattice size) in the physical unit.

\begin{figure}[tbp]
\begin{center}
  \includegraphics[width=7cm, angle=-90]{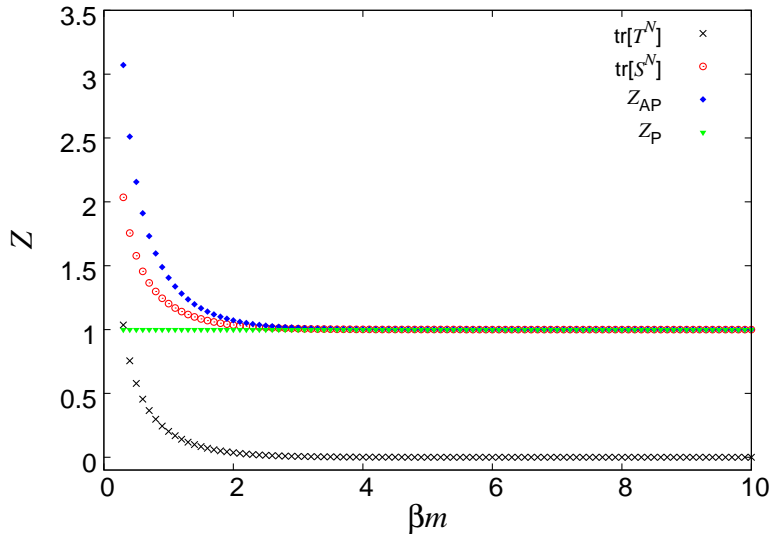}
  \caption{
Partition function against $\beta m$ for $am = 0.01$, $\lambda = 1$, $K=150$ in $\phi^6$ theory. 
 $Z_\m{P}$ (green triangle) and  $Z_\m{AP}$ (blue square) are 
ones with periodic and anti-periodic boundary conditions, respectively.
The traces of $T^N$ (black cross) and $S^N$ (red circle) are also presented. 
}
\label{Z_n3}
\end{center}
\end{figure}

Figure \ref{Z_n3} shows the partition function $Z_P$ against $\beta m$ 
at fixed $am = 0.01$, $K=150$. 
At first,  using the Gauss-Hermite quadrature without rescaling $\phi$ explained in section \ref{improvement},
we use  (\ref{zt}) with the transfer matrices (\ref{naiveT})
 and  (\ref{naiveS}) to compute $Z_P$.
We also plot $Z_{AP}$, which is the partition function with anti-periodic boundary condition, 
evaluated from $Z_\m{AP}={\rm tr}(T^N)+{\rm tr}(S^N)$.
Since one exact supersymmetry leads to the correct Witten index even at finite lattice spacing, 
we find that $Z_P=1$ within the errors of ${\cal O}(10^{-9})$ for all $\beta m$.  
This is because $S$ only has the eigenvalue one and the other 
eigenvalues coincide with those of $T$. 
We also confirm that ${\rm tr}(S^N)$ and ${\rm tr}(T^N)$ rapidly converge as $\beta m$ increases. 
So, in the following, we take $\beta m=30$ which is large enough for the purpose of this section 
since the finite $\beta$ effects are of the order of 
$e^{-\beta m}$. 

\begin{figure}[tbp]
\begin{center}
  \includegraphics[width=7cm, angle=-90]{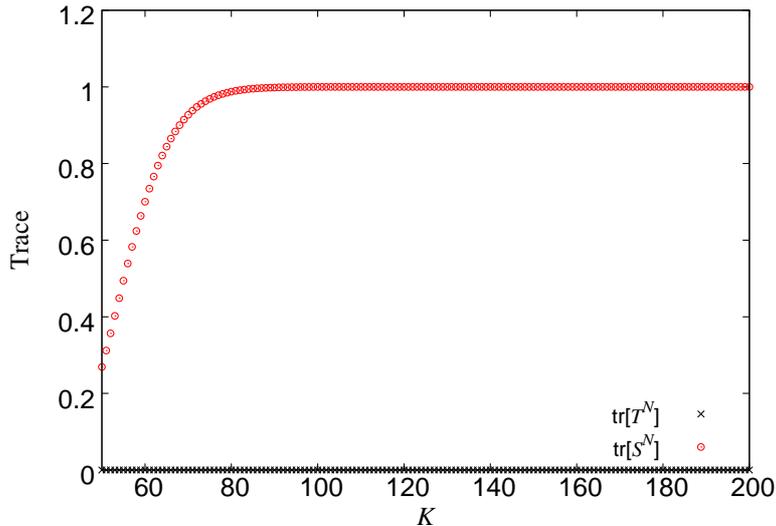}
  \caption{
	Traces of $T^N$ (black cross) and $S^N$ (red circle) with varying $K$, which are
estimated for $N=3000$, $am = 0.01$, $\lambda = 1$ in $\phi^6$-theory. 
}
\label{trace_n3}
\end{center}
\end{figure}

\begin{figure}[tbp]
\begin{center}
  \includegraphics[width=7cm, angle=-90]{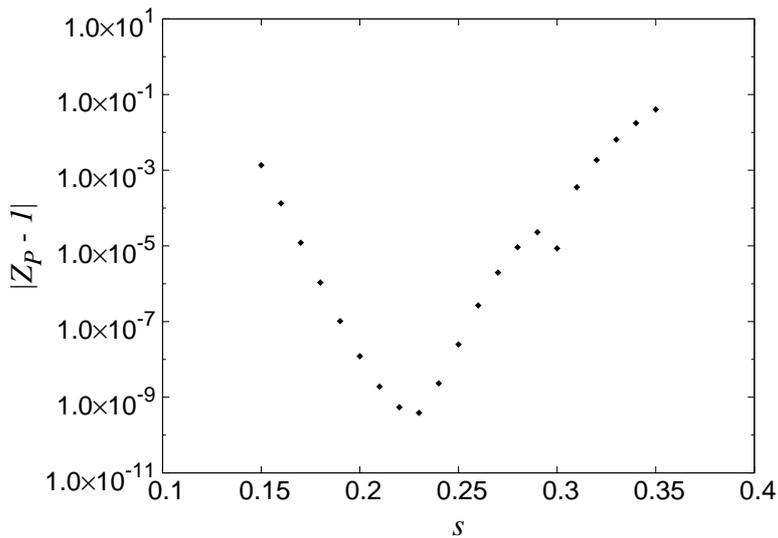}
  \caption{
	$s$-dependence of $|Z_\m{P} - 1|$ for $N=30000$, $am = 0.001$, $\lambda = 1$, $K = 200$.
}
\label{scalepara}
\end{center}
\end{figure}

The Gauss-Hermite quadrature with the scaling 
is employed in the computation.  
We have to choose the number of Gauss-Hermite points $K$ and the scale parameter $s$ for each $ma$.
Although the approximation is expected to be improved as $K$ increases, 
we find that $K \simeq 150-200$ is large enough to show the convergence of results,
for instance, as seen in Figure \ref{trace_n3}. 
For fixed $K$, we optimize $s$ such that the relative error $\delta=|Z_P-1|$ is smaller than $O(10^{-9})$.
Figure \ref{scalepara} is an example of calculation for determining $s$. 
In this case, we have $\delta \lesssim {\cal O} (10^{-9})$ for  $ 0.21  \lesssim s  \lesssim 0.24$.

Table \ref{para_table} shows the parameters used to compute 
the energy eigenvalues and correlators of $\phi^6$ theory, 
chosen by the procedures above.  
We concentrate on the case of 
\be{
\lambda = 1,\qquad \beta m =30.
}
This value of $\lambda$ has been used in the previous Monte-Carlo studies. 
The continuum limit is achieved by taking $ma \rightarrow 0$ for fixed $\beta m$.
In the table, we basically choose $N$ such that the lattice spacings 
are given in equal intervals as $ma = (\beta m) /N $. 
 \footnote{
 In Table \ref{para_table}, 
$ma$ is shown in round off: {\it e.g.} $ma=0.01898\cdots$ for $N=1580$ is written as $ma=0.019$.
 }
Same as the free theory, the scale parameter $s$ approaches zero as $ma$ decreases.
Although we could not find $s$ that realizes $Z_P=1$ within errors of ${\cal O}(10^{-9})$ 
for $ma<0.001$ within $K \le 200$, the lattice spacings are smaller than those used in the Monte-Carlo studies. 
\begin{table}
\begin{center}
\begin{tabular}{cccc} 
\hline\hline
$am$ & s & $N$ & $K$ \\
\hline
0.020&
0.88&
1500&150\\
0.019&
0.86&
1580&\\
0.018&
0.84&
1670&\\
0.017&
0.82&
1760&\\
0.016&
0.80&
1880&\\
0.015&
0.78&
2000&\\
0.014&
0.76&
2140&\\
0.013&
0.74&
2310&$\cdot$\\
0.012&
0.72&
2500&$\cdot$\\
0.011&
0.70&
2730&$\cdot$\\
0.010&
0.68&
3000&\\
0.009&
0.62&
3330&\\
0.008&
0.60&
3750&\\
0.007&
0.55&
4290&\\
0.006&
0.5&
5000&\\
0.005&
0.45&
6000&150\\
0.004&
0.44&
7500&170\\
0.003&
0.38&
10000&170\\
0.002&
0.32&
15000&170\\
0.001&
0.23&
30000&200\\
\hline\hline
\end{tabular}
\end{center}
\caption{
  Parameters used in the computations of $\phi^6$ theory. 
}
\label{para_table}
\end{table}

The energy eigenvalues can be read directly from the transfer matrices $S$ and $T$. 
As already mentioned in section \ref{lat_theory}, $S$ and $T$ correspond 
to two Hamiltonians in SUSY QM, $H_B$ and $H_F$.   
We can read the eigenvalues of bosonic states from $S={\rm exp}(-a H_B)$ 
and ones of fermionic states from $T={\rm exp}(-a H_F)$, 
and obtain $E^{B}_i$ and $E^F_i$  by diagonalizing the transfer matrices $S$ and $T$.
We find that the lattice results satisfy (\ref{EBandEF}) within the relative errors of ${\cal O}(10^{-9})$ 
in all parameters (finite lattice spacings), 
at least, for five smallest eigenvalues.
This is because the one supersymmetry remains exactly on the lattice. 

\begin{figure}[tbp]
\begin{center}
  \includegraphics[width=8cm, angle=-90]{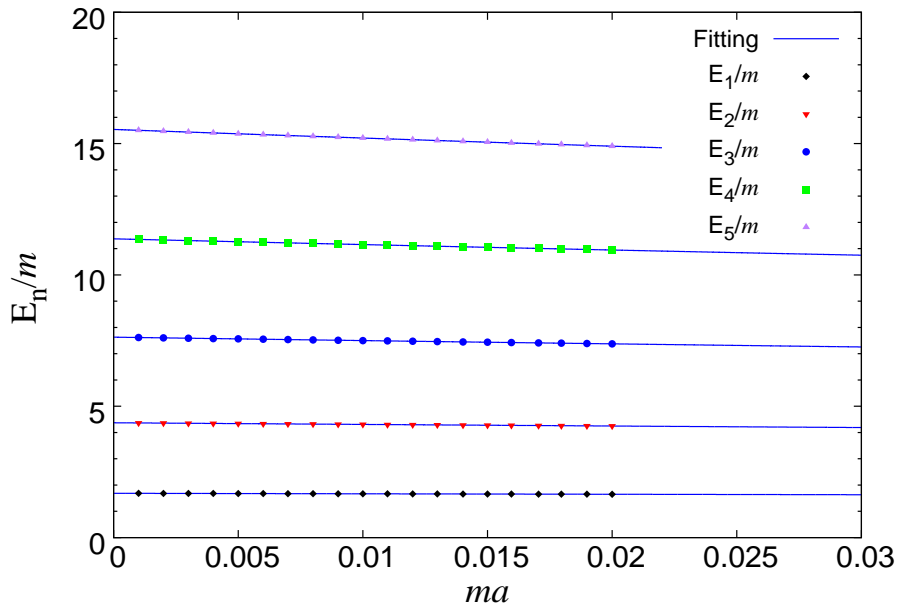}
  \caption{
	Energy spectra against the lattice spacing for $\phi^6$-theory.  
The lowest five eigenvalues are estimated for each $ma$ at fixed $\lambda = 1$, $\beta m = 30$. 
}
\label{extrap_effmass_n3_from_trans}
\end{center}
\end{figure}

Figure \ref{extrap_effmass_n3_from_trans} shows the five smallest non-zero energy eigenvalues
for several lattice spacings. 
To extrapolate the values to the continuum limit, we use a fit formula, 
\be{
E/m=a_0+a_1 (ma) +a_2 (ma)^2.
}
Note that $a_0$ is the value of $E/m$ at continuum limit.
Table \ref{ev_n3} shows the fit results. 
The main results are obtained from a fit for ten data points at small lattice spacings, 
while the errors are given by the largest difference between the main fit and other fits; 
one for twenty points, and fits using cubic polynomial for ten and twenty points.
As explained above, $E_i \equiv E_i^B=E^F_i$ and $E^B_0=0$ 
within the relative errors of $O(10^{-9})$ for all  lattice spacings. 
So the values given in this table are common for $E_i^B$ and $E^F_i$  because 
the errors of the extrapolation are larger than $O(10^{-9})$. 
The extrapolated values of $E/m$ can be determined in very high precision, within the errors less than $0.001$\%.

\begin{table}
\begin{center}
\begin{tabular}{cccccc} 
\hline\hline
&$E_1/m$ & $E_2/m$ & $E_3/m$ & $E_4/m$ & $E_5/m$\\
$a_0$&1.686497(3) & 4.37180(2) & 7.63091(4) & 11.3748(1) & 15.5396(2)\\
$a_1$&-1.897(2)      & -6.417(9)    & -13.29(2)   & -22.39(5)    & -33.69(8)\\
$a_2$&3.0(2)           & 12(1)          & 29(3)          & 55(5)          & 91(9)\\
\hline\hline
\end{tabular}
\end{center}
\caption{
  Energy eigenvalues of $\phi^6$ theory. 
The lowest five eigenvalues are obtained by using a fit, $E/m=a_0+a_1 (ma) +a_2 (ma)^2$,
with $a_0$ the values of $E/m$ at the continuum limit.
}
\label{ev_n3}
\end{table}

\begin{figure}[tbp]
\begin{center}
  \includegraphics[width=7cm, angle=-90]{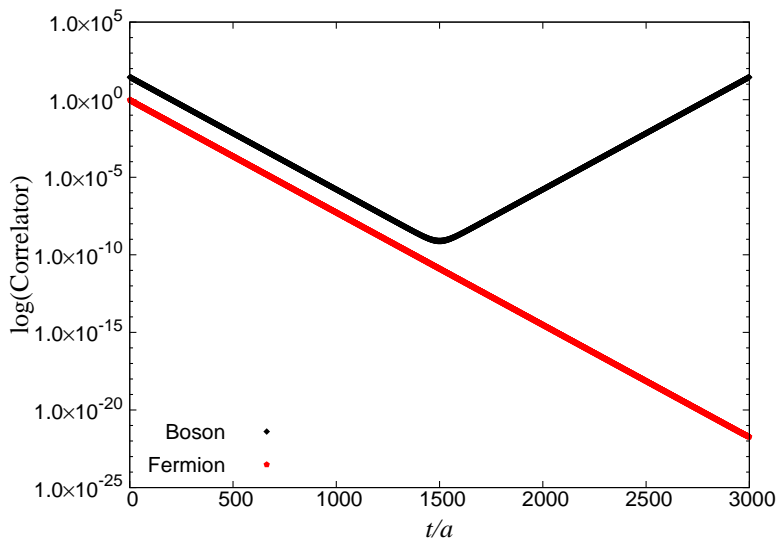}
  \caption{
	Boson and fermion correlators
for $am = 0.01$, $\lambda = 1$ in $\phi^6$ theory.
The results of boson and fermion look like curves, but actually they are black and red points 
plotted for all $t$.
}
\label{cor_mass_n3}
\end{center}
\end{figure}

In Figure \ref{cor_mass_n3}, the correlation functions $\langle\phi_N\phi_t \rangle$ 
and $\langle\psi_t\overline{\psi}_N\rangle$ are shown 
as the black points and the red ones, respectively, in the case of $am=0.01$. 
Those points look like two curves since they are estimated for all $t$ without statistical errors. 
The correlators clearly show the exponential damping, 
and the slope is expected to be $E_1/m$. 
One can obtain any correlators in the same manner as $\langle\phi_N\phi_t \rangle$ 
and $\langle\psi_t\overline{\psi}_N\rangle$, namely, as very clear signals 
since there are no statistical uncertainties in our method. 

Although the energy eigenvalues can be read from the transfer matrices and were already shown above, 
let us apply the standard techniques to extract the effective masses from the correlation functions
for the purpose of making a comparison our approach and Monte-Carlo method.

\begin{figure}[tbp]
\begin{center}
  \includegraphics[width=7cm, angle=-90]{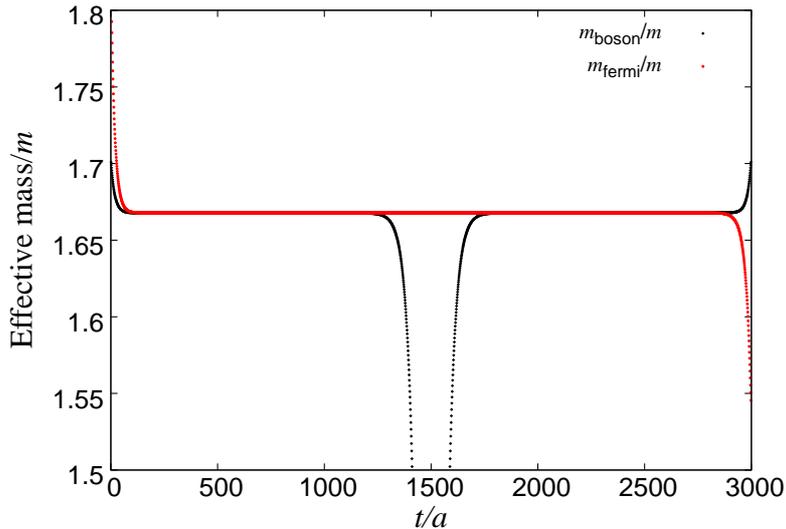}
  \caption{
	Effective masses $am_\m{boson}(t)$ and $am_\m{fermi}(t)$ for $N = 3000$, $am = 0.01$, $\lambda = 1$.
}
\label{eff_mass_n3}
\end{center}
\end{figure}

\begin{figure}[tbp]
\begin{center}
  \includegraphics[width=7cm, angle=-90]{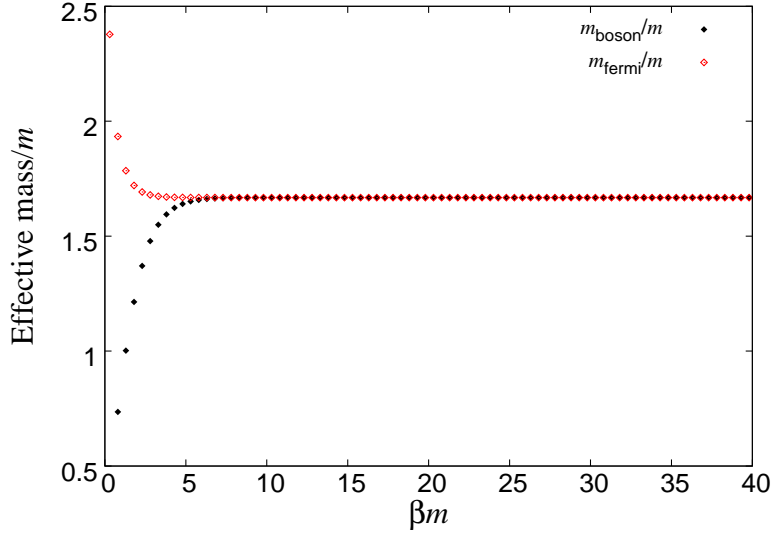}
  \caption{
	Finite volume effect of effective masses.
    We plot $am_\m{boson}$  and $am_\m{fermi}$ estimated at $t=[N/5]$ against $\beta m$, 
    which are measured from the correlators for $am = 0.01$, $\lambda = 1$, $K=150$.
}
\label{latdep_effmass_n3}
\end{center}
\end{figure}

Figure \ref{eff_mass_n3} shows the effective masses defined by
\beqnn{
&& 
m_\m{boson}(t) \equiv \bigg|
\log{\frac{\langle\phi_N\phi_t\rangle}{\langle\phi_N\phi_{t+1}\rangle}} \bigg|,
\\
&& 
m_\m{fermi}(t) \equiv 
\bigg|
\log{\frac{\langle\psi_t\overline{\psi}_N\rangle}{\langle\psi_{t+1}\overline{\psi}_N\rangle}}\bigg|,
}
which are estimated for $ma=0.01$ ($N=3000$).
The signals are also very clear without any statistical errors in comparison with the Monte-Carlo results.
They are not constants 
since the high energy modes contribute near $t=0$ and the finite size effects are seen around $t=N/2$.
The degenerated plateaus are seen for $0 \ll t \ll  N/2$.  
%
Estimating the effective masses at $t \simeq N/5 = 600$,  we have  
$m_\m{fermi}/m=1.6678215773620(2)$ and $m_\m{boson}/m=1.667821577(1)$. 
The errors are estimated by varying $t$ from $500$ to $700$.
This degeneracy is realized in the accuracy of nine digits. 
In Figure \ref{latdep_effmass_n3}, the finite size effect is shown 
 by changing $\beta m$ with the other parameters fixed. 
 The masses are estimated at $t=[N/5]$. \footnote{$[x]$ denotes Gauss symbol, which is the greatest integer less than $x$.}
 We find that $\beta m=30$ is large enough to obtain $m_{\rm eff}$ having no finite size effects. 

\begin{figure}[tbp]
\begin{center}
  \includegraphics[width=8cm, angle=-90]{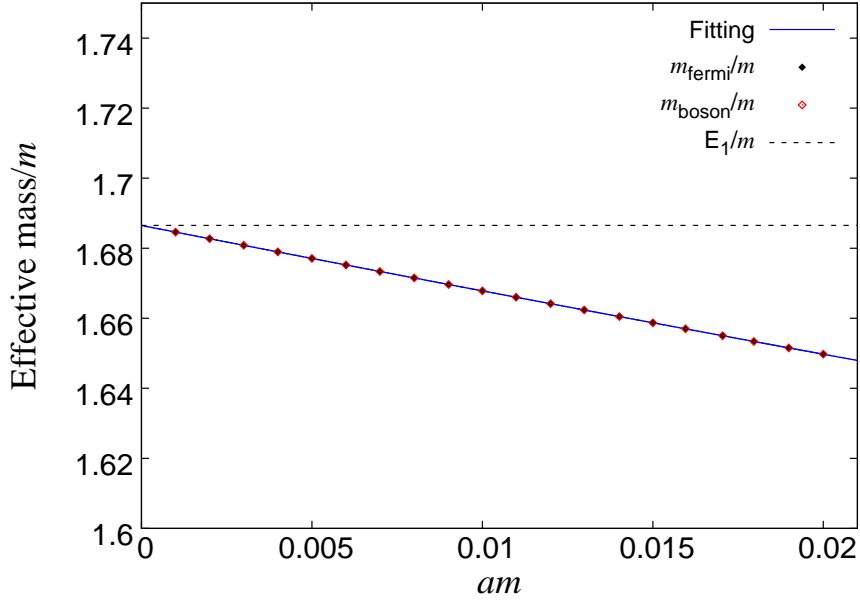}
  \caption{
	Continuum limit of effective masses. $am_\m{boson}$ and $am_\m{fermi}$, which are estimated at $t=N/5$
for $\lambda = 1$, $\beta m = 30$, are plotted.
}
\label{extrap_effmass_n3}
\end{center}
\end{figure}

In Figure \ref{extrap_effmass_n3}, an extrapolation to the continuum limit are shown for the effective masses  estimated at $t=[N/5]$.
We simply use the polynomials $f(x) = a_0 + a_1x + a_2x^2$ as a fit function. 
The systematic errors of the fit are estimated in the same manner as the extrapolation of energy eigenvalues. 
The resultant effective mass is $m_\m{eff}/m = 1.686497(3)$ which is completely same as $E_1/m$ 
shown in Table \ref{ev_n3}.

Finally we show the numerical result of the SUSY Ward identity in the present lattice model.
The SUSY Ward identity is used to show 
whether the broken supersymmetry is restored as a symmetry in the continuum limit.
Consider the lattice supersymmetry transformation of $\langle \overline{\psi}_N \phi_t - \psi_t \phi_N \rangle $. 
Then one find that 
\be{ 
\langle \delta (\overline{\psi}_N \phi_t - \psi_t \phi_N)  \rangle
= \epsilon  {\cal R}^{(1)}(t) + \overline{\epsilon}  {\cal R}^{(2)}(t),  
}
where
\beqnn{
&& {\cal R}^{(1)}(t) = 
-\langle \overline{\psi}_N \psi_t \rangle
-\langle (\nabla_- \phi_N + W(\phi_N) ) \phi_t \rangle,
\\
&&  {\cal R}^{(2)}(t) = 
-\langle \overline{\psi}_N \psi_t \rangle
+\langle  \phi_N (\nabla_+ \phi_t + W(\phi_t) ) \rangle.
}
We can test whether full ${\cal N}=2$ supersymmetry are restored in the continuum limit 
by examining ${\cal R}^{(1)}(t)$  and ${\cal R}^{(2)}(t)$ 
because $\langle \delta (\overline{\psi}_N \phi_t - \psi_t \phi_N)  \rangle$ vanishes in the continuum theory.

Figure \ref{wardn3} shows numerical results of SUSY Ward identity
for free theory ($\lambda = 0$) and interacting case ($\lambda = 1$).
In the free theory, we observe that ${\cal R}^{(1)}(t)={\cal R}^{(2)}(t)=0$ within the error of ${\cal O}(10^{-9})$ for all $t$ since full ${\cal N}=2$ supersymmetry remain on the lattice. 
For interacting case,  although ${\cal R}^{(1)}(t)$ vanishes for all $t$, ${\cal R}^{(2)}(t)$ does not.
We found that it vanishes for $1 \ll t/a \ll N $, 
and  SUSY breaking effect does not survive at a scale lower than the cut-off.  
Thus we find that full ${\cal N}=2$ supersymmetry are restored in the continuum limit even for interacting case.
%
%
%
%
\begin{figure}[tbp]
\begin{center}
  \includegraphics[width=5cm, angle=-90]{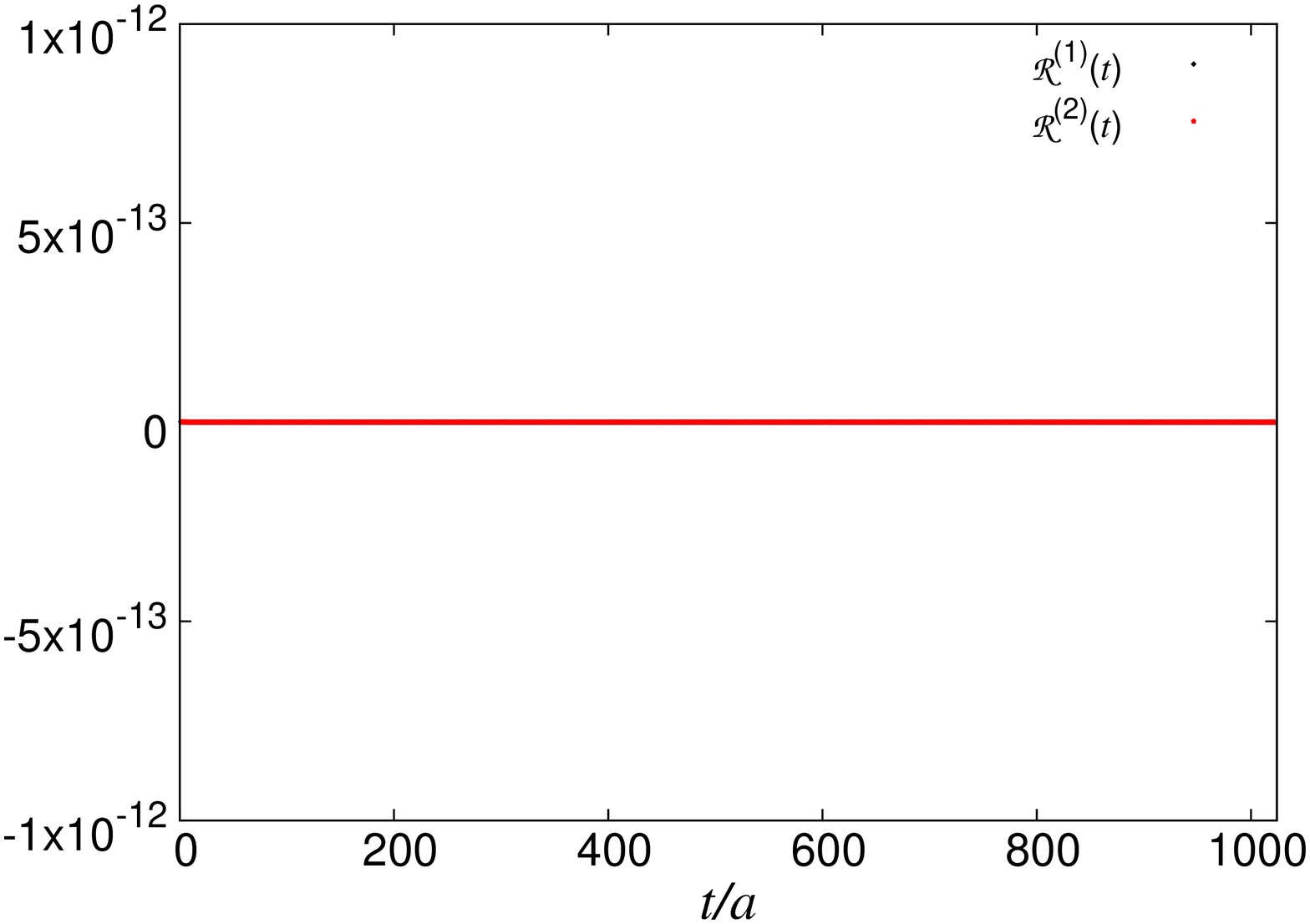}
  \includegraphics[width=5cm, angle=-90]{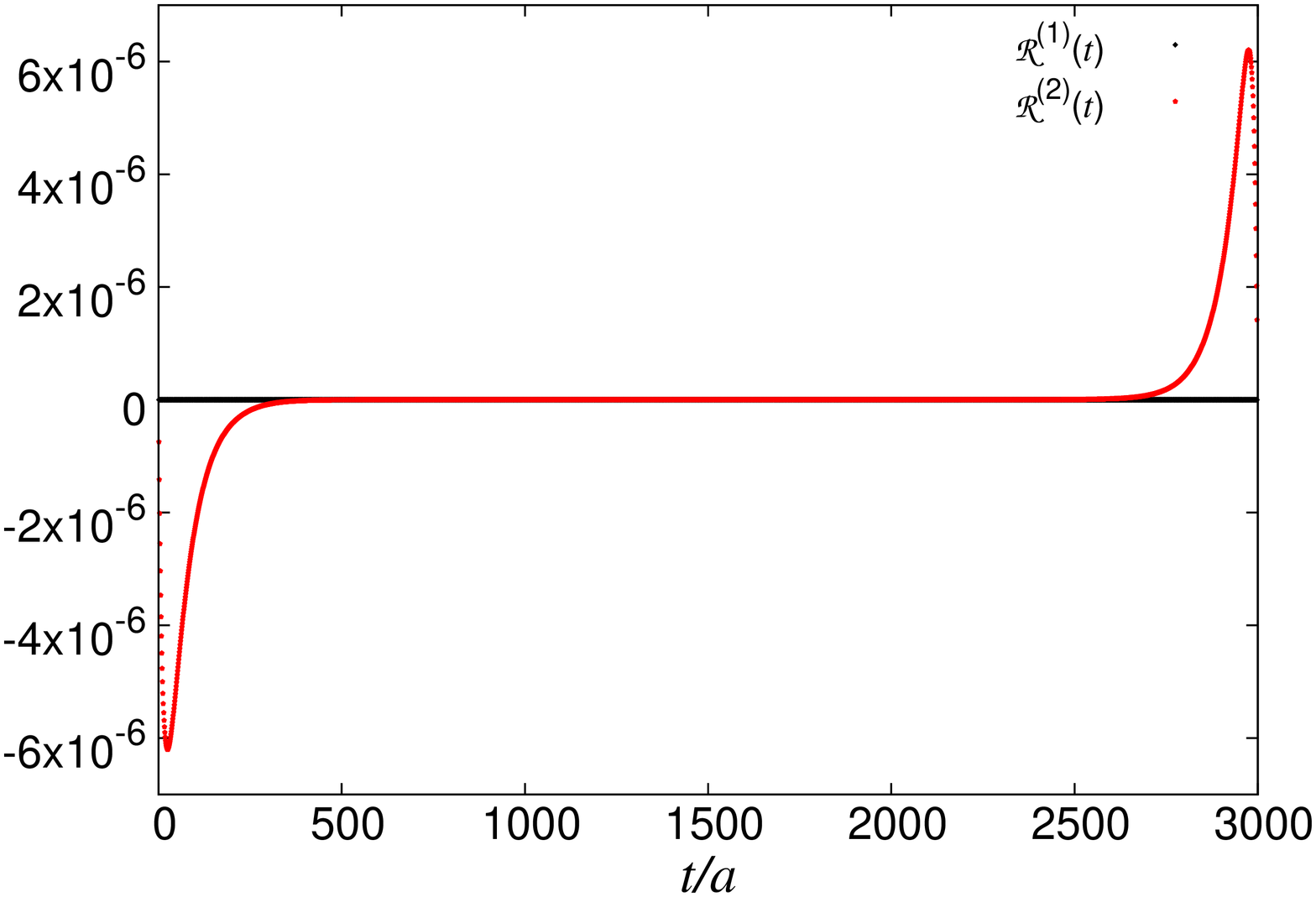}
  \caption{
  SUSY Ward identity in $\phi^6$ theory. 
Left figure shows the results in the free theory ($am = 0.4$ and $\lambda = 0$),
 and Right figure shows an interacting case ($am = 0.01$ and  $\lambda = 1$). 
}
\label{wardn3}
\end{center}
\end{figure}

\subsection{Scarf I potential}
The Scarf I  model is characterized by the superpotential,
\be{
W(\phi)
=
\lambda \alpha \m{tan}(\alpha \phi),
}
where $\alpha$ is a parameter with the mass dimension $1/2$ and $\lambda$ is the dimensionless coupling.
This potential is one of SIPs, and the energy spectra are exactly solved as
\be{
\frac{E_n}{\alpha^2} = \frac{(2n\lambda + n^2)}{2}.
}
We now take 
\be{
\lambda = 10, \qquad \beta \alpha^2 =5,
} 
to test our method in strong coupling region $\lambda \gg 1$. 
One find that the finite size effect  is small enough by taking $\beta \alpha^2 =5$,
since it is of the order of $e^{- \beta E_1}$ and $\beta E_1  \simeq 10 \beta \alpha^2$.

Table \ref{para_table_Scarf} shows our parameters used in the computations of Scarf I model. 
We determine them using the same prescription as that of $\phi^6$ theory.
We also use the Gauss-Legendre quadrature which 
is more effective than Gauss-Hermite method 
for small lattice spacings 
in the sense that $Z_P=1$ for smaller $K$ by choosing the scale parameter $s$.
\begin{table}
\begin{center}
\begin{tabular}{ccccc} 
\hline\hline
$\alpha^2a$  & s & $N$ & $K$ &Quadrature\\
\hline
0.00110&
0.35&
4550&150&Gauss-Hermite\\
0.00105&
0.33&
4760& &\\
0.00100&
0.33&
5000& &\\
0.00095&
0.30&
5260& &\\
0.00090&
0.30&
5560& &\\
0.00085&
0.30&
5880& &\\
0.00080&
0.30&
6250& $\cdot$&\\
0.00075&
0.29&
6670& $\cdot$&\\
0.00070&
0.29&
7140& $\cdot$&\\
0.00065&
0.28&
7690& &\\
0.00060&
0.275&
8330& &\\
0.00055&
0.275&
9100& &\\
0.00050&
0.27&
10000& &\\
0.00045&
0.27&
11100& 150&\\
0.00040&
0.013&
12500&320 &Gauss-Legendre\\
0.00035&
0.012&
14300&340 &\\
0.00030&
0.012&
16700&340 &\\
0.00025&
0.011&
20000&420 &\\
0.00020&
0.010&
25000&440 &\\
0.00015&
0.008&
33300&520 &\\
0.00010&
0.007&
50000&600 &\\
\hline\hline
\end{tabular}
\end{center}
\caption{
  Parameters in Scarf I model.
}
\label{para_table_Scarf}
\end{table}

\begin{figure}[tbp]
\begin{center}
  \includegraphics[width=8cm, angle=-90]{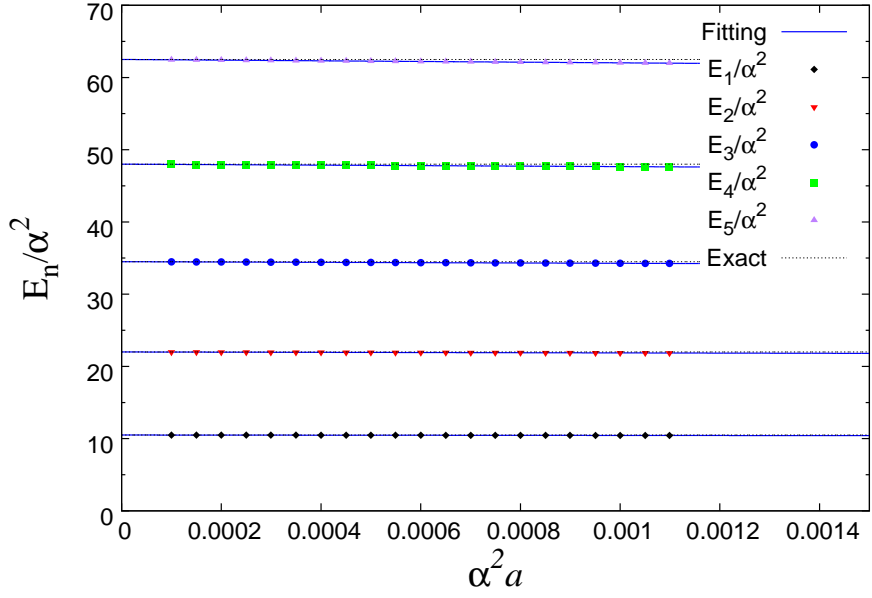}
  \caption{
	Energy spectra against the lattice spacing for Scarf I model.  
The lowest five eigenvalues are estimated for each $\alpha^2 a$  at fixed $\lambda = 10$, $\beta \alpha^2 = 5$. 
}
\label{evs}
\end{center}
\end{figure}

In Figure \ref{evs}, the energy spectra obtained by our method are shown.
Same as $\phi^6$-theory, 
the eigenvalues of $T$ and $S$ satisfy (\ref{EBandEF}) within the errors of ${\cal O}(10^{-9})$.
Table \ref{Spect_from_trans} shows the results of extrapolation to the continuum limit. 
We use a linear function to fit five data points at small lattice sizes, 
and the errors are estimated from the maximum of differences 
between the fit and ones for ten points and using quadratic polynomial. 
The obtained $a_0$ coincide with the exact solution within the small errors less than $0.001\%$. 
\begin{table}
\begin{center}
\begin{tabular}{cccccc} 
\hline\hline
&$E_1/\alpha^2$ & $E_2/\alpha^2$ & $E_3/\alpha^2$ & $E_4/\alpha^2$ & $E_5/\alpha^2$\\
\hline
Exact
&10.5 & 22 & 35 & 48 & 62.5\\
$a_0$&10.49998(2) & 21.99996(5) & 34.4999(1) & 47.9999(2) & 62.4998(2)\\
$a_1$&-57.7(1) & -131.9(3) & -224.0(5) & -335.5(8) & -468.0(12)\\
\hline\hline
\end{tabular}
\end{center}
\caption{
Energy eigenvalues of Scarf I model. 
The lowest five eigenvalues are obtained by using a fit, $E/m=a_0+a_1 (ma)$,
with $a_0$ the values of $E/m$ at the continuum limit.
}
\label{Spect_from_trans}
\end{table}

\begin{figure}[tbp]
\begin{center}
  \includegraphics[width=8cm, angle=-90]{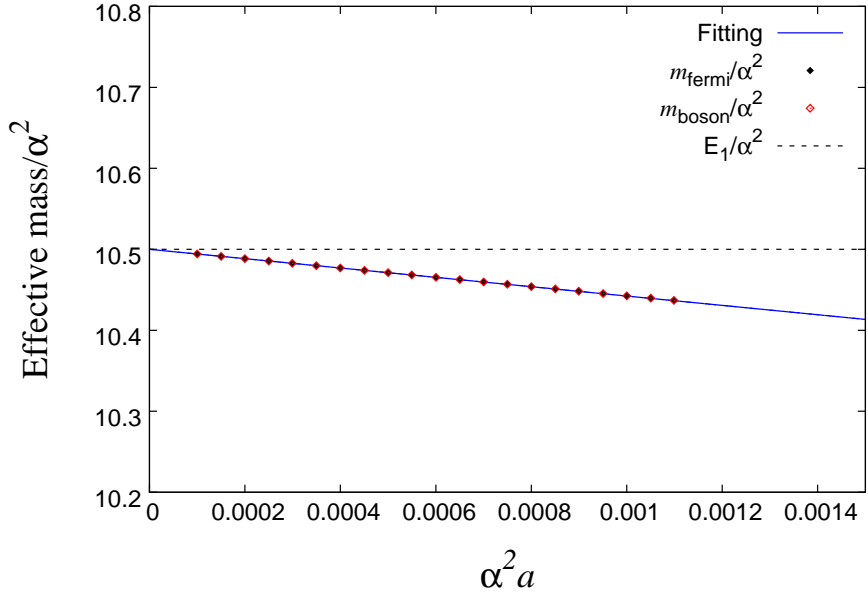}
  \caption{
  Continuum limit of effective masses. $am_\m{boson}$ and $am_\m{fermi}$ are estimated at $t=[N/5]$
for $\lambda = 10$, $\beta \alpha^2 = 5$.
}
\label{extrap_effmass_Scarf}
\end{center}
\end{figure}

We also estimate the effective masses from correlators $\langle\phi_N\phi_t \rangle$ 
and $\langle\psi_N\overline{\psi}_t\rangle$ 
by the standard techniques used in the Monte-Carlo studies.
The detailed procedures were already explained in $\phi^6$-theory.
Figure \ref{extrap_effmass_Scarf} shows the extrapolation of effective masses to the continuum limit.
We employ  $f(x)=a_0 +a_1x$ as a fit formula. The main fit and error estimation are performed 
in the same manner as energy eigenvalues in Table \ref{Spect_from_trans}.  
Since the values at the continuum limit also reproduce the exact solutions with accuracy of $ 99.999\%$, 
we can say that the correlators are evaluated in high precision by our method.

SUSY Ward identity is shown in Figure \ref{wardn3}. 
${\cal R}^{(1)}(t)$ vanishes for all $t$ thanks to the exact  supersymmetry, while
${\cal R}^{(2)}(t)$ vanishes for $1 \ll t \ll N$ as with the case of $\phi^6$-theory. 
Since two identities hold at  a scale lower than the cut-off, full supersymmetry is restored in the continuum limit.
In this way, one can confirm the supersymmetry restoration with high precision, compared to the Monte-Carlo method. 
\begin{figure}[tbp]
\begin{center}
  \includegraphics[width=8cm, angle=-90]{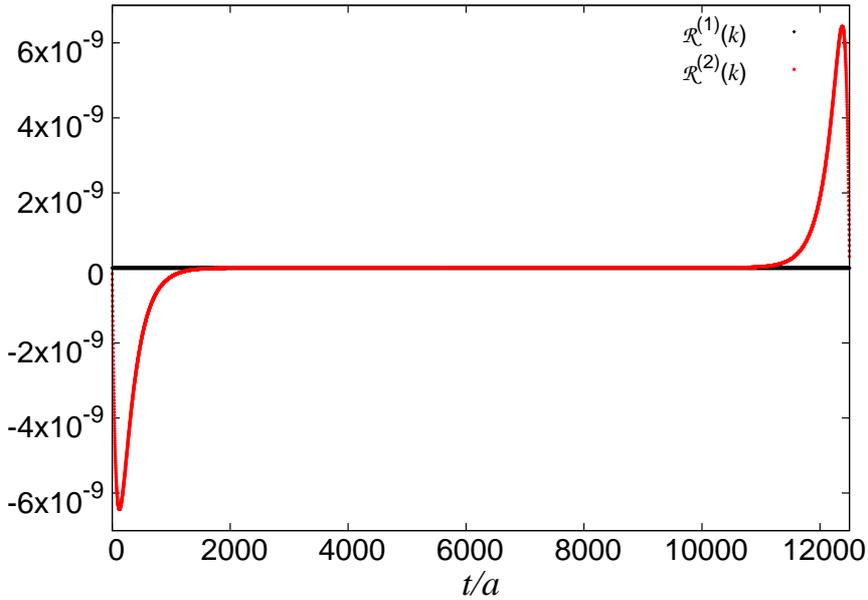}
  \caption{
SUSY Ward identity in Scarf I model, which are estimated for $\lambda = 10$ and $\beta \alpha^2=5$. 
}
\label{wardn3}
\end{center}
\end{figure}

\section{Conclusion and discussion} 
\label{sec:conclusion}

We have proposed a computational method based on the
transfer matrix representation of lattice quantum mechanics.
The numerical quadrature has been used to define the transfer matrices as finite dimensional ones, 
and Gaussian quadrature with rescaling variable went well  
even for interacting cases. We have tested our method for cubic and Scarf I potentials, 
and found that the energy eigenvalues and correlators can be evaluated 
in high precision within reasonable number of discretized points $K$.

Compared to Monte-Carlo method, stochastic processes are not needed
to estimate the expectation values, 
and the sign problem does not exist in this method. 
The systematic error from finite-$K$ effect is well-controlled 
by taking larger $K$ and tuning the scale parameter $s$.
The computational cost grows up with the logarithm of lattice size. 
Thus one can perform the computations at very small lattice spacings by taking large lattice sizes. 
Our method is useful for improving lattice SUSY models since
one can examine the SUSY Ward identity 
and determine the cut-off dependence of physical quantities precisely.

Although we demonstrate our method in  a specific lattice model of SUSY QM,
the main idea of this paper is not limited to that case.
Also, for non-SUSY models, a finite dimensional transfer matrix is defined by the numerical quadrature, 
and one can improve the method by rescaling variables. 
The scale parameter $s$ is chosen such that the Witten index reproduces the correct value 
for the present lattice SUSY model.
However, a method of tuning $s$ is unknown for general models 
and remains an open question. 

The  higher dimensional theories could be studied by extending our method
since the transfer matrix approach is related with not only TNR but also 
worldlines and worldsheets representation 
of $SU(N)$ gauge theory with fermions \cite{deForcrand:2009dh,Marchis:2017oqi,Gattringer:2017hhn}.
These kinds of extensions are applicable to higher dimensional models, 
and the techniques established in this paper could be useful.

%


\acknowledgments

We would like to thank 
Yoshinobu Kuramashi, Yoshifumi Nakamura, Shinji Takeda, Yuya Shimizu,
Yusuke Yoshimura, Hikaru Kawauchi, and Ryo Sakai for valuable comments
 on TNR formulations which are closely related with this study. 
D.K also thank Naoya Ukita for encouraging our study.
This work is supported by JSPS KAKENHI Grant Numbers JP16K05328 
and the MEXT-Supported Program for the Strategic Research Foundation at Private Universities Topological Science (Grant No. S1511006).

\bibliography{Refs}
\bibliographystyle{JHEP}

\end{document}